\documentclass{article}

\usepackage[english]{babel}
\usepackage{authblk}

\usepackage[letterpaper,top=2cm,bottom=2cm,left=3cm,right=3cm,marginparwidth=1.75cm]{geometry}

\usepackage{amsmath}
\usepackage{graphicx}
\usepackage[colorlinks=true, allcolors=blue]{hyperref}

\title{Local distortions as a source of piezoelectric/stiffness decoupling in B-doped AlScN}
\author[1]{Laszlo Wolf}
\author[1]{Geoff L. Brennecka}
\author[1]{Vladan Stevanov\'{c}}
\affil[1]{Department of Metallurgical and Materials Engineering, Colorado School of Mines, Golden CO, 80204, USA}
\date{\vspace{-5ex}}

\begin{document}
\maketitle

\begin{abstract}
We present a first-principles analysis of the wurtzite pseudo-ternary (Al,Sc,B)N to elucidate the structural origin of a decoupling between stiffness $C_{33}$ and piezoelectric response $e_{33}$ upon boron incorporation, using DFT-relaxed 100-atom special quasirandom structures across a broad composition range. Pair distribution function analysis reveals interstitial threefold-coordinated boron atoms that have displaced from the tetrahedral cation site. Direct structural analysis establishes their preferential orientation along the $c$-axis and identifies a scandium-activated creation mechanism. The vertical coordination asymmetry of each cation is quantified through a site-specific axial asymmetry ratio (AAR), showing that boron incorporation progressively symmetrizes the Sc environment. Correlation with Born effective charges demonstrates that this symmetrization is the mechanism behind the piezoelectric enhancement.
\end{abstract}

Aluminum Scandium Nitride (AlScN) alloy has seen a lot of scrutiny in the last 15 years following the discovery of a strong piezoelectric-response enhancement by Akiyama \textit{et al.}\cite{akiyamaAM_2009}. Its improved piezoelectricity over pure AlN put it at the forefront of applications such as radio frequency (RF) filters where surface and bulk acoustic wave (SAW and BAW) filters have been developed around it\cite{wang_APL:2014,umeda_IEEE:2013}. Additionally, its recently discovered ferroelectric behavior\cite{fichtner_JAP:2019,fichtner_APR:2025} brought a second wave of interest to this material with applications in advanced memory devices\cite{qin_NM:2024}.  


The origin of the increase in the piezoelectric response is still debated but the prevailing understanding attributes it to phase competition between wurtzite, rocksalt and a layered hexagonal phase\cite{tasnadiPRL_2010,talley_PRM:2018,chen_PRM:2025}.
This layered hexagonal phase was also pointed out as an intermediate state of the discovered ferroelectric switching behavior of this material\cite{fichtner_JAP:2019,fichtner_APR:2025}.
However it was recently shown\cite{weilee_SA:2024} that at higher alloying content ($>30\%$Sc) this collective switching behavior evolves into an individual switching regime where singular tetrahedra flip consecutively, highlighting the importance of local environments. Similarly, ferroelectricity has been demonstrated in AlBN\cite{hayden_PRM:2021} with a mechanism also based on local bonding and local distortions. 



Co-doping of AlN with Sc and B has been experimentally realized by multiple groups at varying compositions\cite{gremmel_JAP:2025,yousefian_IEEE:2025,saha_FiM:2025,skidmore_JAP:2025} with most of them focusing on the impact of the boron incorporation on ferroelectricity related properties such as so-called wake-up behavior and leakage currents. On the theoretical side, a first-principles study by Jing \textit{et al.}\cite{jing_JAP:2022} predicted that partial replacement of Sc with B in AlScN could simultaneously increase both the piezoelectric $e_{33}$ and stiffness $C_{33}$ coefficients, a surprising departure from the usual negative correlation, as seen in binary AlScN\cite{tasnadiPRL_2010}. That study also noted the potential for B atoms to displace from their tetrahedral cation sites, but did not investigate this phenomenon further.

In this work we use large-supercell DFT calculations to map a broad portion of the composition range of this pseudo-ternary alloy in the Al-rich side. Using a denser composition grid, we aim to obtain further details on the dependence of these properties with respect to the stoichiometry, but most importantly, to connect them to local distortions and coordination changes associated with threefold-coordinated boron atoms.

\begin{figure*}
    \centering
    \includegraphics[width=\linewidth]{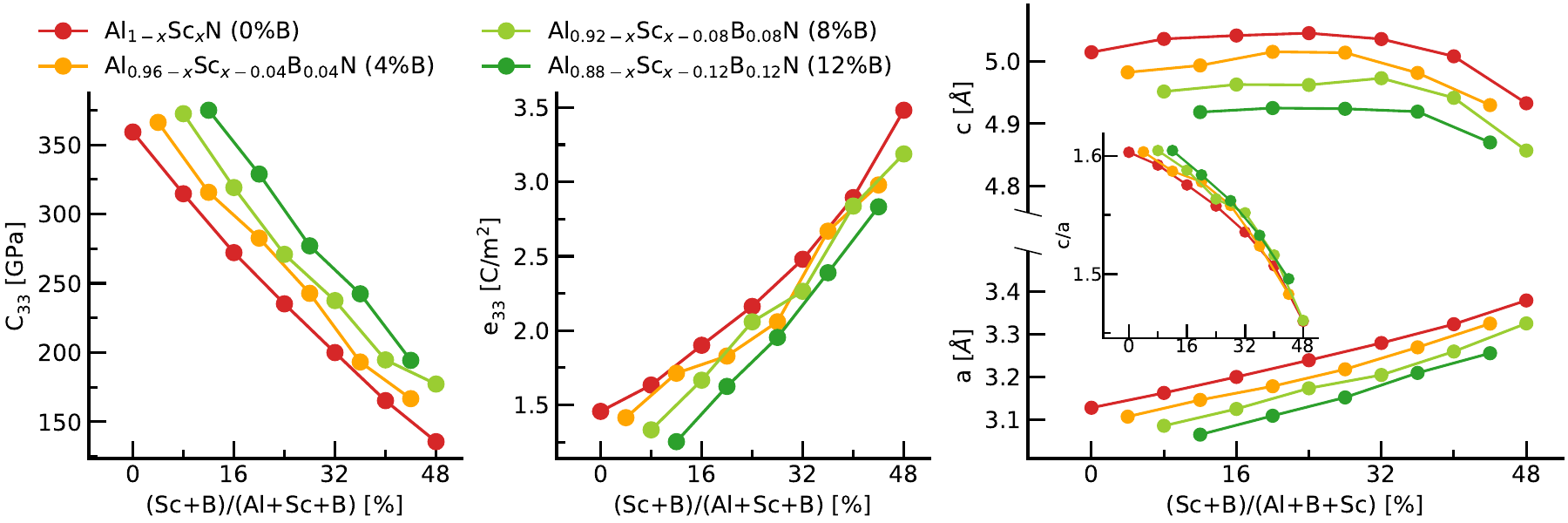}
    \caption{Evolution of properties with varying cation substitution, varying colors relate to the fixed content of boron within those series. Left panel is the 33 component in the Voigt notation of the stiffness tensor, middle panel is the 33 component of the piezoelectric stress tensor and right panel presents lattice parameters of the wurtzite alongside their ratio in the inset}
    \label{fig:1}
\end{figure*}

The wurtzite phase of the (Al,Sc,B)N pseudo-ternary alloy is here represented through 100 atom special quasirandom structures (SQSs)\cite{zungerPRL} with mixed occupation of the cation sublattice by aluminum, scandium and boron atoms while the nitrogen sublattice is left fully occupied. The Monte Carlo procedure to randomize the occupation is performed using the ATAT toolkit\cite{vandewalle_CALPHAD:2009} and by optimizing the supercell shape and the pair-correlation parameter up to the 7th shell and the triple-correlation one up to the 5th shell. Three independent SQS realizations are generated for each composition. The compositions are organized into four series of fixed boron content (0\%, 4\%, 8\%, and 12\% B), with scandium content varied in steps of $\Delta x$ = 8\% at the expense of aluminum, leading to the following series Al$_{100-x}$Sc$_{x}$N, Al$_{96-x}$Sc$_{x-4}$B$_{4}$N, Al$_{92-x}$Sc$_{x-8}$B$_{8}$N, Al$_{88-x}$Sc$_{x-12}$B$_{12}$N. All percentages in this work refer to cation-site fraction, also these four series will mostly be designated by their boron content. Growing high Sc-concentration AlScN alloy in single phase is challenging and strongly dependent on experimental setup\cite{patidar_PRM:2024} and even if higher alloying content have been achieved with B incorporation\cite{saha_FiM:2025}, the present work only considers AlN-rich compositions as the structural and property phenomena of interest are already present in these compositions and exact thermodynamic and phase stability dependence of this system is outside of the scope of this study.

Overall, a total of 60 disordered structures of the pseudo-ternary alloy are obtained alongside the pure AlN, all of which have their cell-shape and ionic position relaxed through DFT where the electron-electron interactions are treated with the PBE exchange-correlation functional\cite{perdewPRL} and the PAW method\cite{blochPRB} as implemented in the VASP computer code\cite{kresse_PRB_1999}. Plane-wave energy cutoff of $340$ eV is used in all calculations alongside automatic generation of the $\Gamma$-centered k-point grid $R_k$ value of 20. The structural relaxation is considered converged when (i) forces on atoms fall below $0.02$ eV$/$\AA, (ii) difference in energy of consecutive steps falls below $10^{-10}$ eV, (iii) total hydrostatic pressure is below $0.5$ kbar. Elastic properties are computed through the finite-differences\cite{le_PRB:2002} method, with an atomic displacement of $0.015$ \AA, the piezoelectric and other related tensors are obtained through density functional perturbation theory\cite{wu_PRB:2005}.

Figure \ref{fig:1} presents calculated stiffness, piezoelectric stress response tensors vertical component and lattice parameters obtained from the various boron content series with the $x$-axis representing the total cation substitution fraction. Each point is an average over the three structures from that specific composition. The $C_{33}$ and $e_{33}$ values are the $c$-axis oriented components of the stiffness and piezoelectric stress response tensors respectively, as given in the Voigt notation. The piezoelectric stress response tensor is built as the sum of the \textit{clamped-ion} and \textit{relaxed-ion} tensors\cite{wu_PRB:2005}. In order to follow experimental procedures but also because SQS supercell vectors do not correspond to the ideal wurtzite unit cell, the lattice parameters are obtained from simulated powder X-ray diffraction (XRD) data by considering the $2\theta$ values of the peaks associated with the $(100)$ and $(002)$ Miller indices. 

The AlScN data presented in Figure \ref{fig:1} (red data) agrees with previous experimental measurements\cite{kurz_JAP:2019,barth_MT:2016,umeda_IEEE:2013,akiyamaAM_2009,lu_APLmat:2018,petrich_2019} within the understood uncertainty of DFT and DFPT for lattice parameters\cite{Zhang_2018} and mechanical properties calculations\cite{Jong_SD:2015Piezo,Jong_SD:2015Elastic}. See supplemental materials for experimental comparison of the piezoelectric strain coefficient $d_{33}$ .



For series containing B, one can analyze the reproduction of the AlBN alloys, here by considering the first point of each curve. Both $c$ and $a$ linearly drop with increasing B content, with $c$ going from $5~$\AA\, to $4.9~$\AA\, at $12$\%B and $a$ from $\sim3.1~$\AA\,
to $\sim3.05~$\AA\, over the same range\cite{hayden_PRM:2021}. 

The decoupling between the $c$-axis stiffness and the $33$ component of the  piezoelectric stress tensor can be seen in the Figure~\ref{fig:1} data when observing the impact of boron incorporation at a fixed alloying content. There is a clear impact on the $C_{33}$ and lattice parameters $c$ and $a$, identifiable by the evenly-spaced curves which shows a linear impact of B incorporation at a fixed AlN content. However, this does not hold for either $e_{33}$ or $c/a$, where the curves converge and the impact of boron incorporation is greatly diminished. 

As an example, $C_{33}$ exhibits a steady increase of about 10\% for each increment of 4\% B incorporation in lieu of Sc, and does so no matter the fixed alloying content of AlN. On the other hand $e_{33}$ initially presents a similar steady decrease with B incorporation which disappears at higher alloying content. At 40\% total cation substitution, the 0\%B, 4\%B and 8\%B series are almost overlapping. These results in effect demonstrate the decoupling of the $c$-axis stiffness and piezoelectric properties of this pseudo-ternary and represents a departure from their usual inversely proportional relationship. This decoupling was previously demonstrated by Jing \textit{et al.}\cite{jing_JAP:2022}. The present work finds a more modest intensity of this effect, notably regarding the $e_{33}$ values. This difference in intensity of the decoupling could originate from a fundamental methodology difference between these works where to represent each composition Jing uses an ensemble of small cells (16 atoms) whereas we use three large cells (100 atoms). This difference in size will reflect in particular in the treatment of local effects that will be less adequately captured in small cells due to stronger self-correlation of atoms.

\begin{figure}
    \centering
    \includegraphics[width=0.5\linewidth]{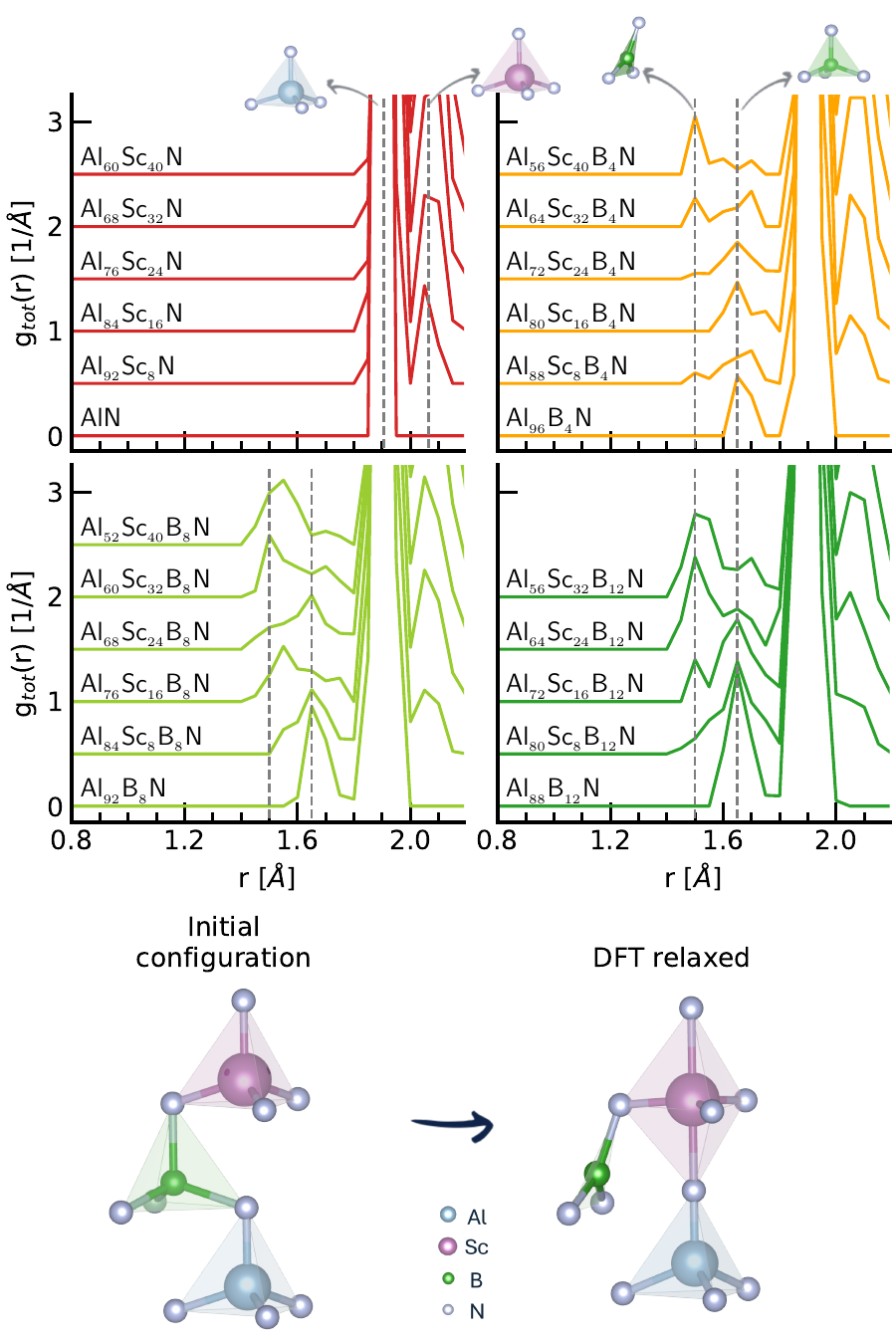}
    \caption{Upper panel presents the low-$r$ portion of the total pair distribution function of the different series, vertical shift is applied with increased Sc content. Lower panel is schematic representation of the transition from tetrahedral to planar configuration of a B atom.}
    \label{fig:2}
\end{figure}

The main local structural effect observed here is the spontaneous relaxation of B atoms that move away from their cation site and now reside in a threefold-coordinated configuration, where they occupy the center of one of the faces of their coordination tetrahedron, as seen in the lower panel of Figure \ref{fig:2}. Evidence of these planar boron atoms is presented in the upper panel through the total pair distribution functions $g_{tot}(r)$ of our different series. The $g_{tot}(r)$ can be considered as a histogram of inter-atomic distances and is computed as a double sum over the atom types $\alpha,\beta$\cite{wolfJAP-2025,keen_JAC:2001},
\begin{equation}\label{eq:pdf}
g_{tot} (r) \, = \, \sum_{\alpha} \, \sum_{\beta} \, \left< \, \frac{1}{4\pi r^2 n c_{\beta}} \, \frac{dN_{\alpha\beta}(r)}{dr} \,\right>_{\alpha},
\end{equation}
with $n$ the number density, $c_{\beta}$ the atomic fraction of the $\beta$ type, $dN_{\alpha\beta}(r)$ represents the number of $\beta$ atoms in a spherical region of the thickness $dr$ at the distance $r$ from a particular
$\alpha$ atom, and the angle brackets $\langle~\rangle_{\alpha}$ denote an average over all $\alpha$ atoms.  

The four subplots of the upper panel of Figure \ref{fig:2} distinguish the different B content series, and within each subplot, a vertical shift is applied with increased Sc content, reproducing the $x$-axis evolution from Figure \ref{fig:1}. In the AlScN data (upper-left panel) the expected Al-N and Sc-N bonds in their tetrahedral coordination can be seen at $1.905~$\AA\, and $2.065~$\AA\, respectively. Most importantly, no shorter inter-atomic distances are present. In B-containing series, the bottom PDF of each subplot corresponds to the AlBN alloy, where a new peak is present at $1.65~$\AA, characteristic of a B-N bond length in its tetrahedral configuration. Upon introduction of Sc, this peak slowly fades and gives way to a lower-$r$ peak at $1.5~$\AA, associated with the shorter B-N bond length characteristic of the planar threefold-coordinated configuration. With greater introduction of Sc, the $1.5~$\AA\, peak grows larger than the $1.65~$\AA\, one, such that in all three B-containing series, planar B become predominant over the conventionally tetrahedral. At the highest Sc content considered for each series, planar B outnumber tetrahedral ones by two to three times (determined from direct coordination analysis). While these peaks in the $g_{tot}$ are a clear and visual indicator of structural changes, the rest of the structural analyses in this paper are done through direct study of the relaxed structures giving more precise information of the orientation of these planar B, as discussed below.

Interestingly, these planar boron atoms show a strong preference for the interstitial sites of the ``vertical" faces of their original tetrahedron rather than for the one in the basal plane of the wurtzite. This behavior notably differs from the BN hexagonal ground-state which features threefold-coordinated boron sheets perpendicular to the $c$-direction. Here, only 6\% of the planar B are in the basal plane while all the others are in the center of one of the ``vertical" faces of their tetrahedron. The origin of this preference can likely be attributed to the larger area (and associated interstitial site) of the vertical faces arising from the elongated $c$-oriented metal-nitrogen bonds characteristic of the wurtzite phase.

The lower panel of Figure \ref{fig:2} illustrates the statistically most common configuration of this structural transition.
The key feature of this transition is an isolated N atom sitting below a Sc, which incorporates it into its coordination shell, approaching a bipyramidal geometry similar to that found in the layered-hexagonal phase. This mechanism accounts for approximately 75\% of planar B instances in our structures. It also explains why no planar B are observed in the AlBN compositions as the transition requires Sc to disturb the structure (increase of $c$) but also to attract and accommodate the soon-to-be-isolated N atom, consistent with the preferred sixfold coordination of scandium in its ScN rocksalt ground state. However these specific Sc atoms are not the only ones exhibiting significant geometry changes, and local distortions are observed throughout our structures.



\begin{figure*}
    \centering
    \includegraphics[width=\linewidth]{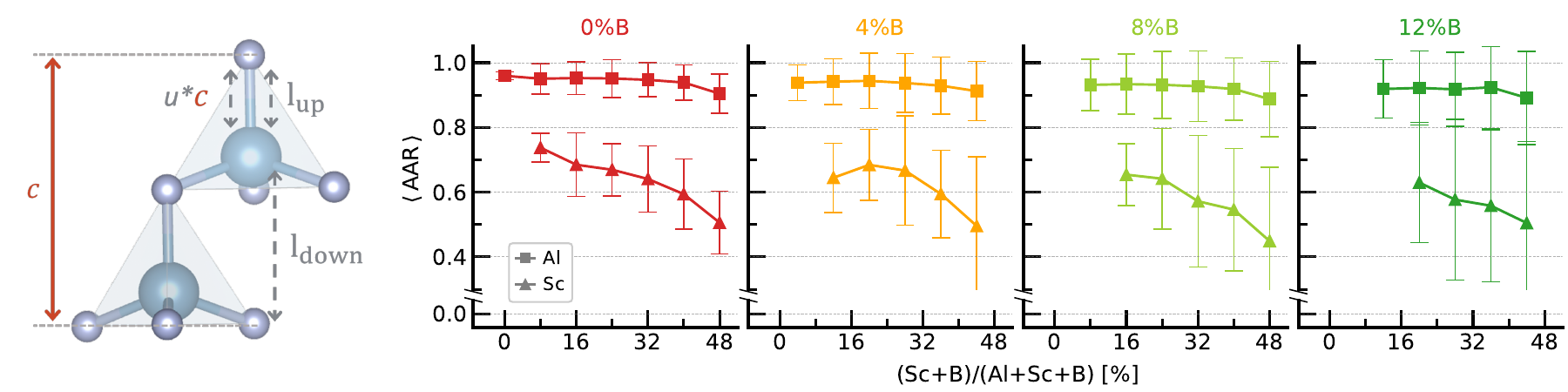}
    \caption{Left panel presents definition of the distances involved in the definition of the axial asymmetry ratio ($\mathrm{AAR}$). Right panels are the averaged $\mathrm{AAR}$ over Sc and Al atoms for each composition. Error bars represent the standard deviation of the corresponding $\mathrm{AAR}$ data set.}
    \label{fig:3}
\end{figure*}

Distortions of the wurtzite lattice are conventionally quantified through the internal parameter $u$ defined from the bond length $l_{up}$ and the $c$ lattice parameter, as seen in the left panel of Figure \ref{fig:3}. This $u$ is well-defined only in a homogeneous structure where cation geometries are consistent, so that $l_{up}$ and $l_{down}$ values are well coupled. The significant local distortions present in the structures studied here decorrelate these two bonds, requiring a metric that accounts for both $l_{up}$ and $l_{down}$ independently, which we define as an atom specific axial asymmetry ratio ($\mathrm{AAR}$),
\begin{equation}
    \mathrm{AAR}\,=\,\frac{l_{down}\,-\,l_{up}}{0.25\,*\,\langle c\rangle}.
    \label{eq:AAR}
\end{equation}
The dividing term corresponds to the ideal wurtzite instance where $u*c=l^{id}_{up}=0.375*c=1-l^{id}_{down}$ such that $l^{id}_{down}-l^{id}_{up}=0.25*c$. Note that due to local distortions we here make explicit the difference between the $c$ distance of Figure \ref{fig:2} and the $\langle c\rangle$ term corresponding to the overall $c$ lattice parameter of the structure, here obtained from simulated XRD data (see Figure \ref{fig:1}), but for simplicity, any mention of the $c$ lattice parameter throughout this paper refers to the overall lattice parameter of the structure. With this metric, a cation presenting an ideal wurtzite vertical asymmetry would have $\mathrm{AAR}=1$ and by contrast, a cation that would present $l_{up}=l_{down}$, as in a bipyramidal configuration, would have $\mathrm{AAR}=0$. For reference, the accepted slight distortion of the wurtzite in pure AlN that gives $u=0.38$\cite{ambacher_JAP_2023} gives $\mathrm{AAR}=0.96$.

\begin{figure}
    \centering
    \includegraphics[width=0.5\linewidth]{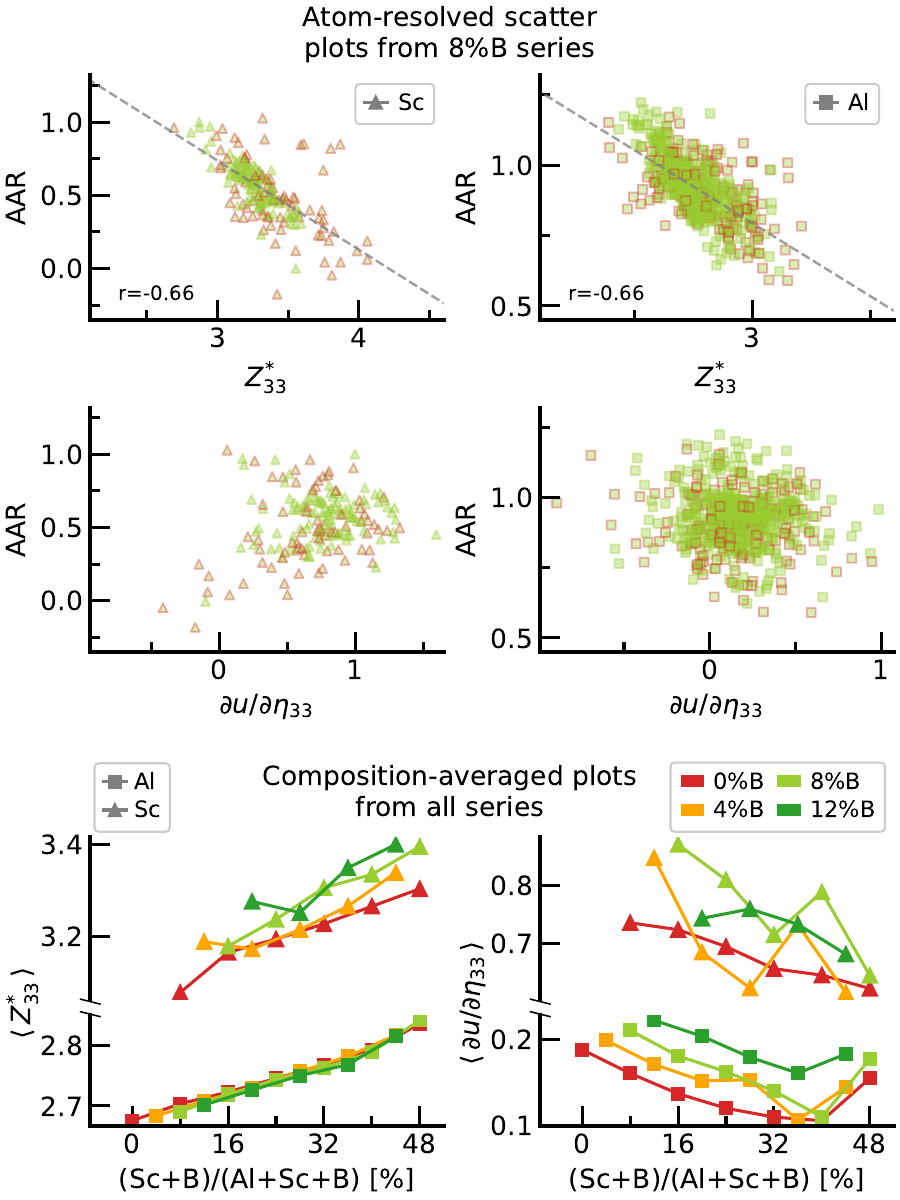}
    \caption{Upper panel shows correlation between the $\mathrm{AAR}$ and both the $Z^{*}_{33}$ and the $\frac{du_{k}}{d\eta_{33}}$ of each individual scandium (triangle) and aluminum (square) atoms of the whole 8\%B series. Red-edged marker are cations that present a planar B in their second shell. Lower panel presents the evolution of $Z^{*}_{33}$ and $\frac{du_{k}}{d\eta_{33}}$ of all series over the alloying range, where each data point is obtained by averaging over all adequate atoms in the structures.}
    \label{fig:4}
\end{figure}

The right panels of Figure \ref{fig:3} present the averaged $\mathrm{AAR}$ over the Al and Sc atoms of all considered structures alongside the associated standard deviation of the values through the errorbars. Overall, the data show that Sc atoms are less tetrahedral (larger distortions) than Al ones. With increasing Sc content this difference grows larger, showing that Sc atoms grow more distorted while Al geometries remain mostly unaffected. Results also agree quantitatively with $u$ values reported by Tasn\'{a}di \textit{et al.}\cite{tasnadiPRL_2010} for the AlScN series, where the decrease in Sc $\mathrm{AAR}$ from $0.72$ to $0.56$ corresponds to an increase of $u$ from $0.41$ to $0.43$. Similarly the Al $\mathrm{AAR}$ stays relatively constant around 0.9 which is $0.38\leq u \leq 0.39$. The impact of boron on these systems can be seen through an overall downward shift of each curve alongside an increase in the spread of each data set, with again these effects impacting the scandium curves with a greater magnitude. A direct interpretation of these results is that boron atoms greatly disturb the homogeneity of the structure (at least in the $c$-direction) and exacerbate local ionic transitions from the polar tetrahedral towards the non-polar bipyramidal geometry. An effect that is amplified for atoms in the vicinity of planar B atoms, as shown in Figure \ref{fig:4}.

To connect these distortions back to the properties of interest, one must consider the decomposition of the piezoelectric stress tensor component $e_{33}$ into clamped-ion and relaxed-ion contributions\cite{wu_PRB:2005}.
\begin{equation}
    e_{33} = e^{clamped}_{33} + \frac{1}{\Omega}\sum_{k}\mathrm{e}Z^{*}_{k,33}\frac{du_{k}}{d\eta_{33}},
    \label{eq:e33}
\end{equation}
where $\Omega$ is the wurtzite unit-cell volume, the sum runs over all atoms of the structure, $\mathrm{e}$ is the electronic charge, $Z^{*}_{k,33}$ is the dynamical Born effective charge (BEC) in units of $\mathrm{e}$ of atom $k$ and $\frac{du_{k}}{d\eta_{33}}$ is the sensitivity of the fractional coordinate in $c$ to a vertical external strain of atom $k$.\footnote{Strictly, these should be written as $e_{z3}$, $Z^{*}_{k,zz}$, $\eta_{3}$ following the full tensor notation, for simplicity, we use the 33 index throughout.} The $C_{33}$ stiffness tensor component has definition that similarly includes $\frac{du}{d\eta_{33}}$. The BEC quantifies the change in polarization with respect to an atomic displacement while the sensitivity quantifies the ease with which an atom displaces under strain. High piezoelectric response therefore requires both a large BEC and high sensitivity, while high stiffness benefits from low sensitivity, which highlights the inherent tension between these two properties that the present decoupling partially resolves.

With these definitions in mind, Figure \ref{fig:4} presents correlation between $\mathrm{AAR}$ and both $Z^{*}_{33}$ and $\frac{du_{k}}{d\eta_{33}}$ of each individual scandium and aluminum atoms. Only the atoms from the 8\%B series are presented here to conserve space but the analysis is similar for the other series (see Supplementary Materials for complete data). Cations whose second coordination contained a boron atom that underwent the tetrahedral to planar transition upon relaxation are highlighted with a red border. Linear regression fits of the $Z^{*}_{33}$ data sets are shown as grey dashed lines with their Pearson correlation coefficients given in the lower-left corner of both panel ($p<10^{-20}$ in both cases). The lower panel presents the evolution of the averaged $Z^{*}_{33}$ and $\frac{du_{k}}{d\eta_{33}}$ for both aluminum and scandium atoms.

From the $Z_{33}^{*}$ scatter data, a clear negative correlation can be seen with $\mathrm{AAR}$ with a comparable slope for Sc and Al ($-0.61$ and $-0.52$ respectively). This correlation shows that, as the cation symmetrizes its geometry in the $c$-direction, its polarization becomes more sensitive to further displacements, consistent with the physical picture of the symmetric bipyramidal geometry as a polarization inflection point between the two polar tetrahedral configurations\cite{weilee_SA:2024,calderon_science:2023}. This behavior is also observed for nitrogen atoms (see Supplementary Materials). On the other hand $\frac{du_{k}}{d\eta_{33}}$ does not show any clear relation to $\mathrm{AAR}$ such that the $c$-direction symmetry around a cation does not uniformly impact its positional sensitivity to strain. Only Sc atoms with very low $\mathrm{AAR}$ values show a significant reduction in sensitivity, demonstrating that moderate vertical symmetry variations have little effect on sensitivity and that significant changes require the full bipyramidal transition involving formation of a new bond which, as previously shown by Jing \textit{et al.}\cite{jing_JAP:2022} has an important impact on this quantity. 

The study by Jing \textit{et al.} reaches a complementary conclusion by showing that the $\frac{du_{k}}{d\eta_{33}}$ correlates with bond strength descriptors derived from integrated crystal orbital Hamilton population (ICOHP) for both in-plane and vertical bonds around each N atom. Such that stronger bonds lead to lower sensitivity. The present findings are consistent with and extend this picture as the $\mathrm{AAR}$, which only measures vertical geometric asymmetry, does not predict sensitivity, which is instead governed mostly by in-plane bond strengths.\cite{jing_JAP:2022}

This analysis is confirmed in the lower panel of Figure \ref{fig:4} where the $Z_{33}^{*}$ curves mirror the inverse of their $\mathrm{AAR}$ counterparts from Figure \ref{fig:3} where the Al curves are almost superimposed while the Sc ones present a shift with an increase in boron content. Meanwhile a link between the sensitivity curves and the $\mathrm{AAR}$ is less evident even if one finds again that the disordered geometries of scandium atoms are reflected here while the aluminum curves present a more ordered and consistent evolution. 

The mechanical properties of this pseudo-ternary alloy and their dependencies on boron incorporation can be understood through two dominant contributions. The $C_{33}$ is maintained by short and intrinsically stiff B-N bonds while $e_{33}$ is enhanced by the scandium atoms whose vertical coordination geometry symmetrizes due to the overall structural inhomogeneity of the structure. Both of these effects are accentuated by the scandium-activated transition of boron atoms into a predominantly $c$-oriented planar threefold-coordinated geometry. A geometry in which the sensitivity of the bonded nitrogen atom remains comparatively low, preserving $C_{33}$. The activating scandium, by accommodating the isolated N, further approaches a vertically symmetric bipyramidal, thereby increasing its $Z^{*}_{33}$ and contributing to the enhancement of $e_{33}$. This mechanism however exhibits diminishing returns. If a proper bipyramidal geometry is achieved, the Sc sensitivity is suppressed, eliminating its piezoelectric contribution while allowing it to positively contribute to $C_{33}$. This suggests an optimal B content that maximally symmetrizes Sc vertical geometry without driving too many sites into full fivefold coordination. 


In summary, we show that the increase of boron content in the wurtzite (Al,Sc,B)N pseudo-ternary alloy significantly disturbs local geometries throughout the structure and brings an overall lowering of the vertical asymmetry of Sc atoms' coordination environment. This evolution is in part helped by the transition of a majority of B atoms from their fourfold coordinated cation site toward a threefold coordinated interstitial site, further shortening the B-N bonds compared to Al-N or Sc-N ones and thus increasing distortions of the lattice. This lowering of the vertical asymmetry, measured by the specifically introduced axial asymmetry ratio ($\mathrm{AAR}$), is strongly correlated to an increase of the Born effective charge $Z^{*}_{33}$ of the associated atom. 
In contrast the $\mathrm{AAR}$ shows no significant correlation to the strain sensitivity $\frac{du_{k}}{d\eta_{33}}$ which is linked to $C_{33}$ and is instead controlled by in-plane environment\cite{jing_JAP:2022}
Together these results identify two structurally orthogonal mechanisms through which boron incorporation enhances $e_{33}$ while maintaining $C_{33}$, offering a design principle for wurtzite piezoelectric alloys with independently optimized stiffness and electromechanical response.

\bibliographystyle{abbrv}
\bibliography{biblio}

\newpage

\section*{Supplementary Material}\label{sec:SM}

\begin{figure*}[!ht]
    \centering
    \includegraphics[width=0.5\linewidth]{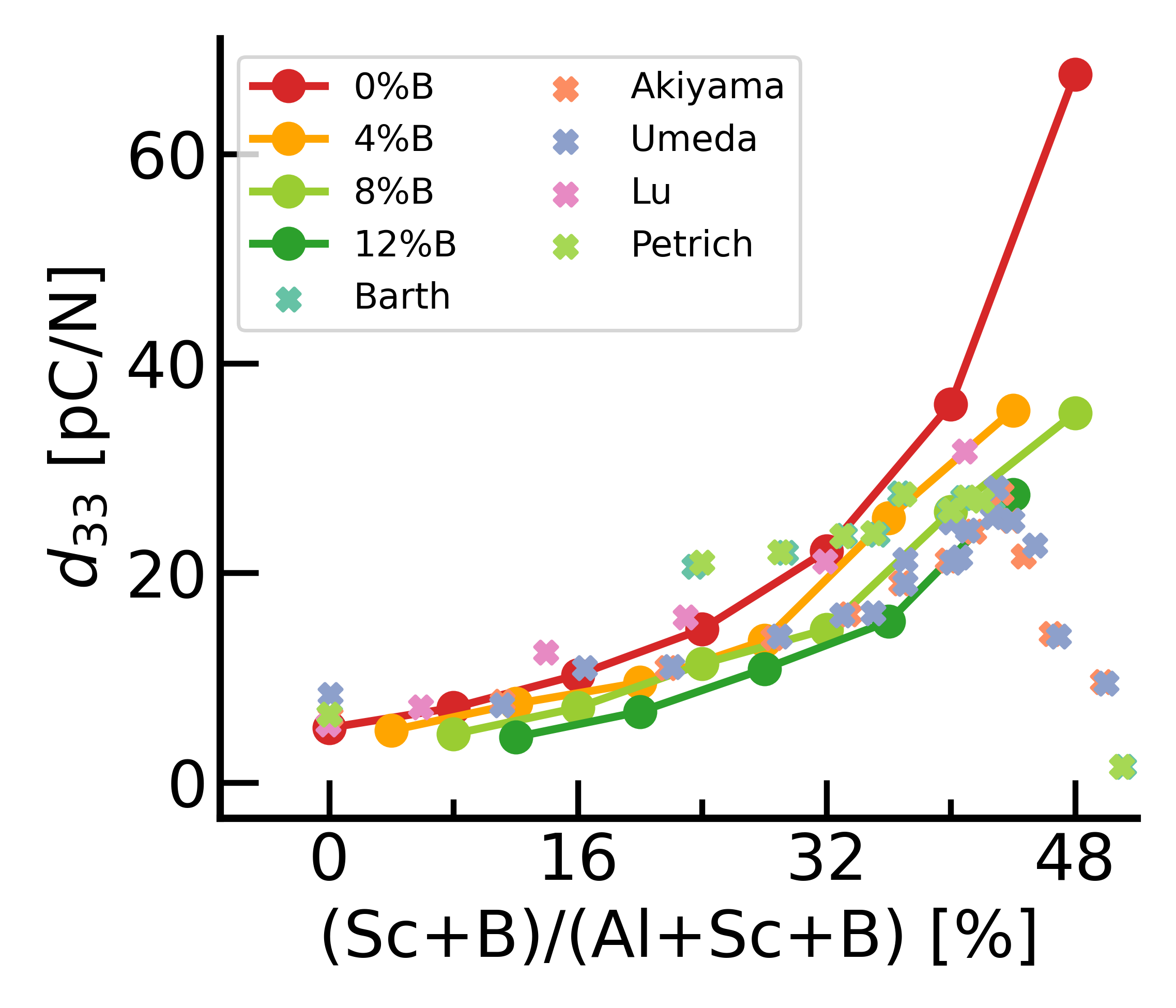}
    \caption{Evolution of the piezoelectric strain response tensor component $d_{33}$ with respect to composition for all four series. Colored cross markers are experimental data from AlScN samples\cite{barth_MT:2016,akiyamaAM_2009,umeda_IEEE:2013,lu_APLmat:2018,petrich_2019}}
    \label{fig:SI-d33}
\end{figure*}

\begin{figure*}[!ht]
    \centering
    \includegraphics[width=\linewidth]{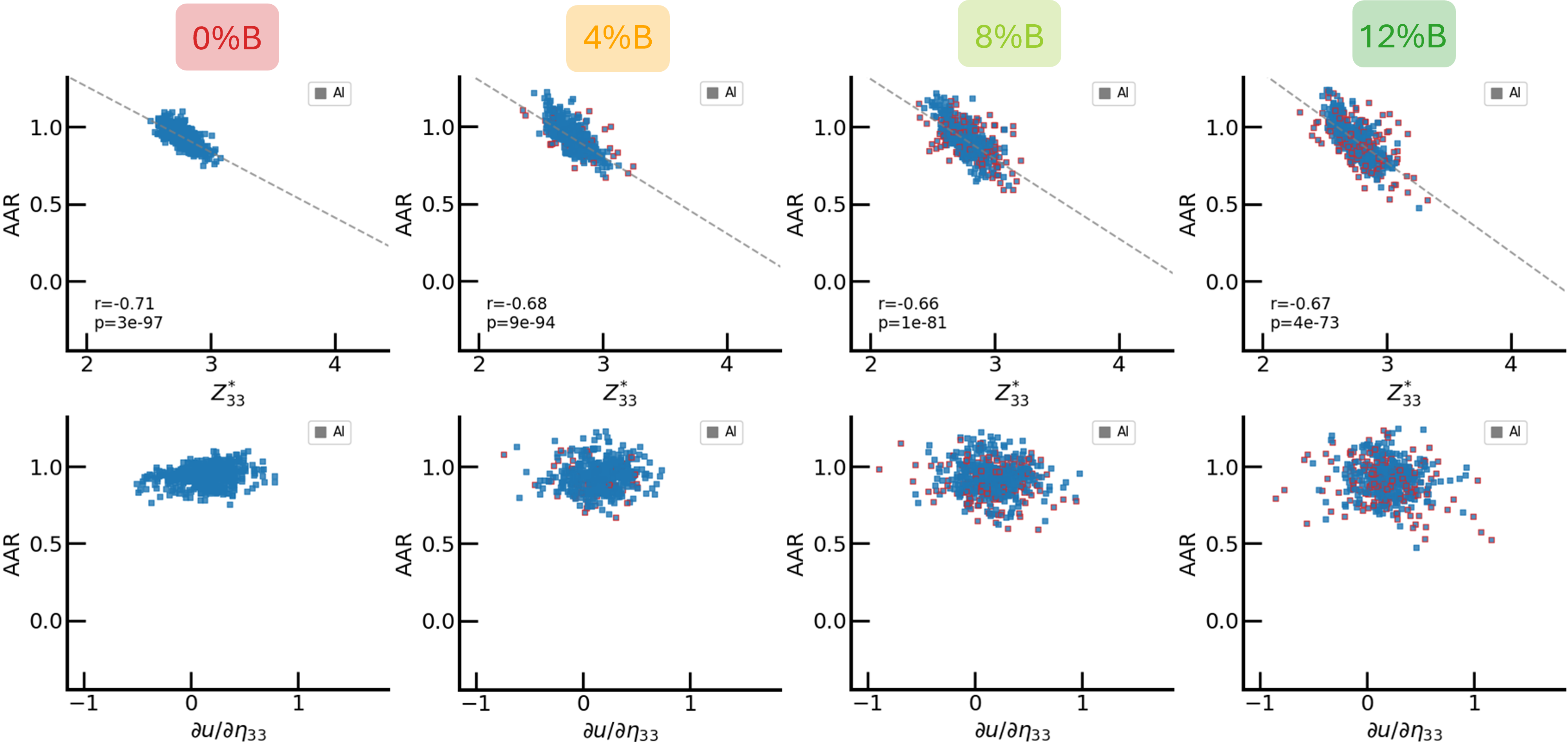}
    \caption{Scatter plot correlation of the axial asymmetry ratio $\mathrm{AAR}$ with the Born effective charges $Z_{33}^{*}$ and the sensitivity of the fractional coordinate in $c$ to a vertical
external strain of individual aluminum atom in all structures from all four series. Red-edged marker are Al atoms that present a planar B in their second shell.}
    \label{fig:SI-AlZ33}
\end{figure*}

\begin{figure*}[!ht]
    \centering
    \includegraphics[width=\linewidth]{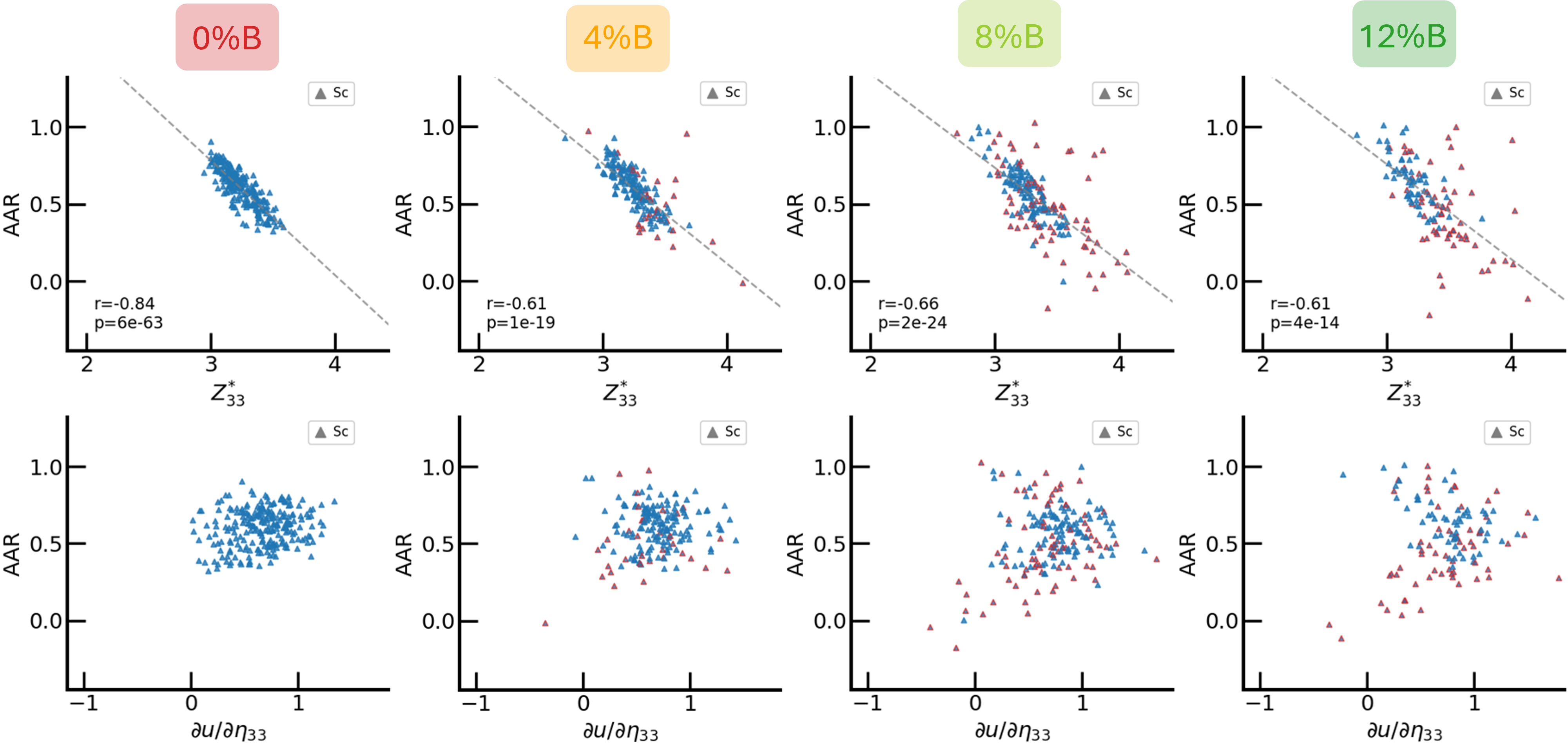}
    \caption{Scatter plot correlation of the axial asymmetry ratio $\mathrm{AAR}$ with the Born effective charges $Z_{33}^{*}$ and the sensitivity of the fractional coordinate in $c$ to a vertical
external strain of individual scandium atom in all structures from all four series. Red-edged marker are Sc atoms that present a planar B in their second shell.}
    \label{fig:SI-ScZ33}
\end{figure*}

\begin{figure*}[!ht]
    \centering
    \includegraphics[width=\linewidth]{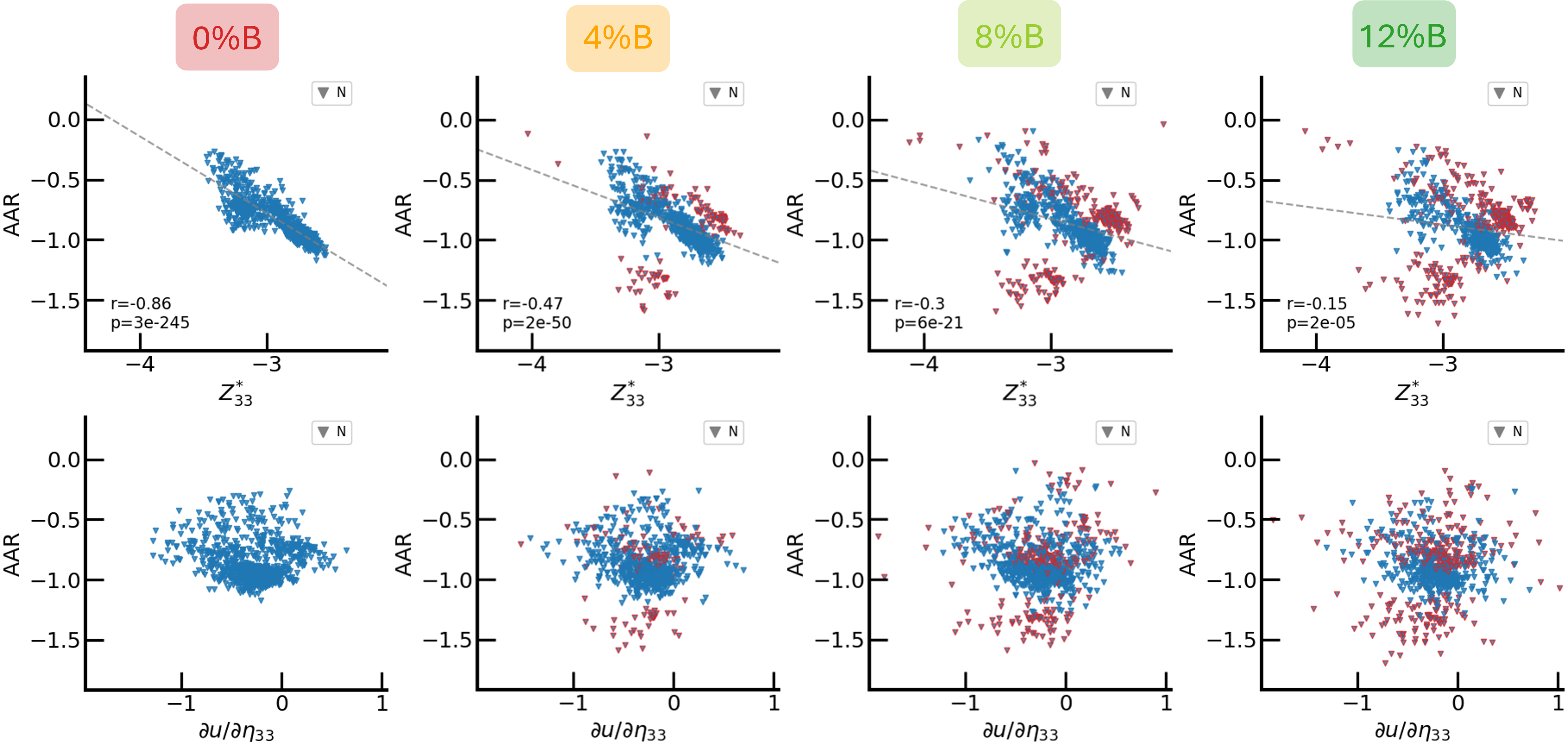}
    \caption{Scatter plot correlation of the axial asymmetry ratio $\mathrm{AAR}$ with the Born effective charges $Z_{33}^{*}$ and the sensitivity of the fractional coordinate in $c$ to a vertical
external strain of individual nitrogen atom in all structures from all four series. Red-edged marker are N atoms that present a planar B in their first shell. Note that the $\mathrm{AAR}$ is defined similarly as for cations by centering on a nitrogen but keeping the same orientation, leading to negative values of $\mathrm{AAR}$}
    \label{fig:SI-NZ33}
\end{figure*}

\end{document}